\documentclass[11pt,epsf]{article}
\usepackage{graphics}
\usepackage{amsmath}
\usepackage{latexsym}
\usepackage{graphics}
\usepackage{rotating}

\makeatletter
\renewcommand{\section}{\@startsection{section}{1}{0in}
        {0.4\baselineskip}{0.1\baselineskip}{\Large\bf}}
\renewcommand{\subsection}{\@startsection{subsection}{2}{0in}
        {0.25\baselineskip}{-\baselineskip}{\large\bf}}
\renewcommand{\subsubsection}{\@startsection{subsubsection}{3}{0in}
        {0.1\baselineskip}{-\baselineskip}{\normalsize\bf}}
\makeatother

\begin{document}

%

\thispagestyle{myheadings}

\markright{OG 2.2.15}
OG 2.2.15 \\

\begin{center}

{\LARGE \bf Test by JANZOS of the Standard Model of Cosmic Ray  
Acceleration
in the COMPTEL/ROSAT Supernova Remnant }
\end{center}

\begin{center}

{\bf F. Abe$^{9}$, I. Bond$^{1,2}$, E. Budding$^{5}$, A. Daniel$^{1}$,
P. Dobcsanyi$^{1}$,  R. Dodd $^{1,3,4}$, H. Fujii$^{14}$, \\
 Z. Fujii$^{9}$,
N. Hayashida$^{11}$, J. Hearnshaw$^{2}$, K. Hibino$^{10}$,  S.  
Kabe$^{14}$, K.Kasahara$^{13}$, \\
P. Kilmartin$^{1,2}$,
S. Laurence$^{1}$, A. Masaike$^{8}$,   Y. Matsubara$^{9}$,  Y.  
Muraki$^{9}$,
 T. Nakamura$^{7}$,  \\
G. Nankivell$^{4}$, M. Nishizawa$^{12}$,
 N. Rattenbury$^{1}$,  M. Reid$^{3}$,  N. Rumsey$^{4}$,  M. Sakata$^{6}$,  
\\
 H. Sato$^{7}$, M. Sekiguchi$^{11}$,
D. Sullivan$^{3}$, M. Teshima$^{11}$,  C. Tsai$^{1}$, \\
 Y. Yamamoto$^{6}$,
 T. Yanagisawa$^{9}$,  P. Yock$^{1}$,  T. Yuda$^{11}$ and  Y.  
Watase$^{14}$ \\
}

{\it $^{1}$University of Auckland, Auckland, NZ, \
$^{2}$University of Canterbury, Christchurch, NZ, \\
$^{3}$Victoria University, Wellington, NZ, \
$^{4}$Carter National Observatory, Wellington, NZ, \\
$^{5}$CIT, Wellington, NZ, \
$^{6}$Konan University, Kobe, Japan, \\
$^{7}$Kyoto University, Kyoto, Japan, \
$^{8}$Fukui University of Technology, Fukui, Japan, \\
$^{9}$STE Laboratory, Nagoya University, Nagoya, Japan, \
$^{10}$Kanagawa University, Yokohama, Japan, \\
$^{11}$ICRR, University of Tokyo, Tanashi, Tokyo, Japan, \\
 $^{12}$National Center for Science Information Systems,Tokyo,
Japan, \\
 $^{13}$Shibaura Institute of Technology, Ohmiya, Japan, \
 $^{14}$KEK Laboratory, Tsukuba, Japan \\
}
\end{center}

\begin{center}
{\large \bf Abstract\\}
\end{center}
\vspace{-0.5ex}

A search for ultra-high energy gamma-rays emitted by the young, nearby  
supernova remnant  that was discovered recently by the COMPTEL and ROSAT  
satellites was made using the JANZOS database for the period 1987-1993. A  
$95 \% $ confidence upper limit on the flux above 100 TeV of  
$3\times10^{-13} cm^{-2} sec^{-1}$ was obtained. This is an order of  
magnitude below the expected flux based on the standard model of cosmic  
ray acceleration in supernova shocks. An optical survey of the region  
that has been commenced is also reported. This uses UK and ESO Schmidt  
plates, and CCD images by a NZ/Japan microlensing group.

\vspace{1ex}

%
\vspace{1ex}
\section{Introduction:}
\label{intro.sec}
The supernova remnant (SNR) that was found recently by the COMPTEL and  
ROSAT satellites (GRO0852-4642/RXJ0852.0-4622)
 (Aschenbach et al. 1998 and  Iyudin et al. 1998) provides an excellent  
opportunity to test the standard model of cosmic ray acceleration in  
supernova shocks. This is because the SNR is very close, $\sim 200$ pc,  
and at an optimal age for cosmic ray acceleration, $\sim$ 680 years.

Cosmic ray acceleration at the SNR is expected to be accompanied by the  
production of gamma-rays through interactions of the cosmic rays with the  
local swept-up interstellar medium. The gamma-rays may be used as a  
tracer of the cosmic rays. The expected flux of gamma-rays at Earth with  
energies greater than 100 TeV is $\sim$few$\times 10^{-12}$ cm$^{-2}  
$sec$^{-1}$ (Naito and Takahara 1994, Drury et al 1994). Such a flux is  
easily detectable with current technology. We have examined the database  
of the JANZOS air shower array at energies $>$100 TeV. This array was  
situated at the appropriate latitude for the COMPTEL/ROSAT SNR and was  
operational from 1987 to 1995 (Bond et al 1988, Allen et al 1993, Allen  
et al 1995). We report here our results for the period 1987-1993.

We also report an optical study of the region using UK and ESO Schmidt  
plates and CCD images obtained by the MOA gravitational microlensing  
collaboration (Abe et al 1996, Alcock et al 1997, Reid et al 1998, Yock  
1998, Abe et al 1999, Yanagisawa et al 1999, Rhie et al 1999, Muraki et  
al 1999).

\vspace{1ex}
\section{Experiment:}
\label{experiment.sec}

The data were obtained with an array of 45 fast timing scintillation  
detectors of area 0.5m$^{2}$ that were deployed from 1987 to 1995 over an  
area 90m $\times 90$m in the Black Birch range in NZ at latitude  
$42^{\circ}$S and altitude 1640m. A detailed description
of the detection system is seen in Bond et al (1988) and Allen et al  
(1993).
It has been found that, for showers with 10$\sim $19 timing measurements,  

 $70 \%$ had the zenith correctly determined to within    $ \pm  
0.90^{\circ}$.
A similar result was obtained in the azimuthal direction. For showers  

with $\geq 20 $ timing measurements the accuracy was found to be better.
This accuracy has been confirmed by observations of the cosmic ray  
shadows
cast by both the sun and the moon (Allen et al 1995).

\vspace{1ex}
\section{Result:}
\label{Results.sec}
The distribution of air shower events recorded by the JANZOS array  close  
to SNR GRO0852 \\ /RXJ0852.0-4622 from 1987 to 1993 is shown in Figure 1.  
 The diameter of the X-ray image of the SNR is about $2^{\circ}$, and the  
bin size is $1^{\circ}\times 1^{\circ}$. The distribution includes air  
showers produced by both cosmic rays and gamma-rays. The former are known  
to be much more numerous, and to have a nearly isotropic arrival  
distribution at Earth, because they are charged particles and deflected  
by the Galactic magnetic field. The observed data are consistent with  
this, the apparent dependence on declination being an air-mass effect.
No clear evidence for an excess of gamma-rays in the direction  


\begin{figure}[h]
\centerline{\includegraphics{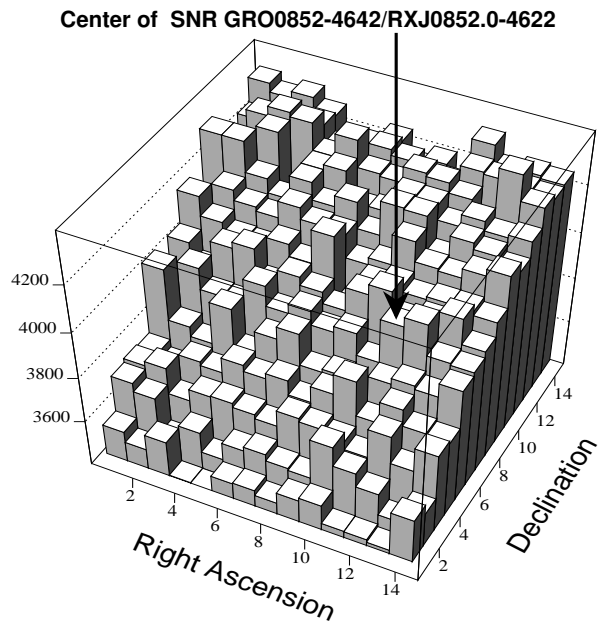}}
\caption{Distribution of air showers detected by the JANZOS array near  
the  COMPTEL/ROSAT SNR (GRO0852-4642/RXJ0852.0-4622). The boundaries are  
at $118.^{\circ}58$ and $138.^{\circ}41$ in right ascension, and at  
$-52.^{\circ}00$ and  $-38.^{\circ}00$ in declination. The bin size is  
$1^{\circ}\times1^{\circ}$. The diameter of the X-ray image of the SNR is  
about $2^{\circ}$. The significance of the excess in the direction of the  
SNR is +0.9 $\sigma$. }
\end{figure}

\newpage
\noindent
of the SNR is seen. To quantify this, the distribution of events in a  
strip of width $2^{\circ}$ in declination and length $23^{\circ}$ in  
right ascension centered on the SNR was examined.
The significance of the excess in the direction of the SNR was found to  
correspond to +0.9 $\sigma$ using Li-Ma statistics (Li and Ma 1983). A  
similar result was obtained using the constant zenith angle technique of  
Acharya et al (1990).

The effective detection time and area were $3 \times10^7 $sec and $1.0  
\times 10^4$ m$^2$ respectively, and the 95$\%$ confidence upper limit on  
the flux of gamma-rays with energies $>$100 TeV is $3 \times10^{-13}  
$cm$^{-2} $s$^{-1}$. 
These values are given in Table 1. The measured upper limit is an order  
of magnitude below the expected flux cited in Section 1.
\vspace{5mm}

Table 1.  Result on the flux of gamma-rays from  SNR  
GRO0852-4642/RXJ0852.0-4622\\
\begin{tabular}{| l | l | l | l | l|}\hline
 Mode energy & Events in SNR & Background in strip &  Excess  (Li-Ma) &  
Flux upper Limit \\
\ \ \ \ \  (TeV) & \ \ \ bin($2^{\circ}\times 3^{\circ}$)   & \ \ \ \  
bin($2^{\circ}\times 20^{\circ}$) & \ \ \ \ \ \ \ ($\sigma$) &  (95$\%$  
conf. level) \\
            &                &                  &                    &   
($\times10^{-13} cm^{-2}s^{-1}$)  \\ \hline
\ \ \ \ \ 100 & \ \ \ \ \ \ 23,622 & \ \ \ \ \ \ 156,464 &  \ \ \ \ \ \ +  
0.9 & \ \ \ \ \ \ \ $3.0  $ \\  \hline
\end{tabular}
\vspace{0.5cm}
\vspace{1ex}

\section{Optical Observations:}

Apparently there is no historical record of the COMPTEL/ROSAT SNR, even  
though a supernova approximately 250 pc away might be expected to  
outshine the full moon. We have commenced an examination of ESO and UK  
Schmidt plates in the R, J, SR and I  passbands, but see no obvious  
nebulosity that matches the ring-like X-ray image of the SNR.

We have also taken several exposures in broad red and blue passbands of  
the region using a wide-field Boller \& Chivens telescope at the Mt John  
University Observatory in NZ, and a large CCD camera of the MOA  
collaboration referred to in the Introduction.
These exposures are currently being stacked
to yield images that are expected to have limiting magnitudes fainter  
than 22.
The image below is a preliminary stacked image in the red passband of a  
region covering $\sim 0.5$ square

\begin{figure}[h]
\centerline{\includegraphics{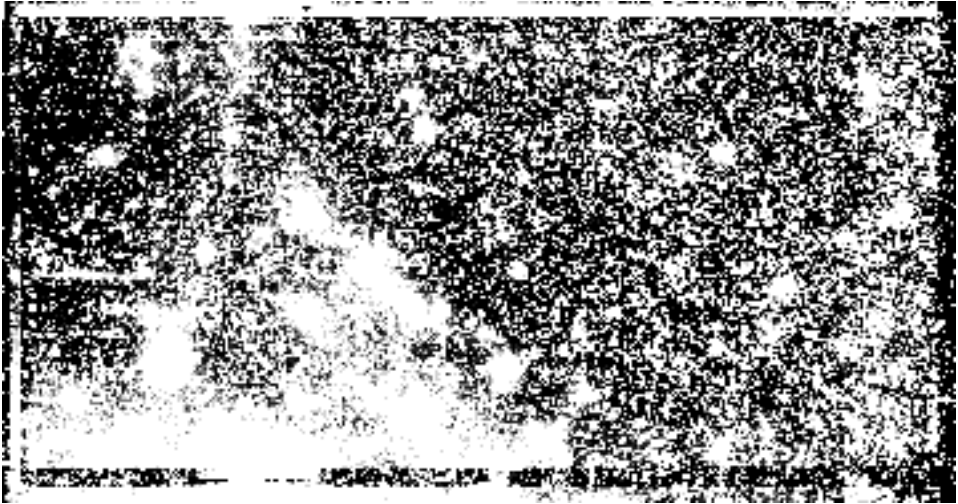}}
\caption{Optical image in red light of 0.5 square degree near
the northern rim of the COMPTEL/ROSAT SNR. The image extends
from  $-46^{\circ}45' $ to $ -47^{\circ}15' $ in declination (2000),
and from 8h 50m to 8h 54m in RA. The black and white intensity
limits are approximately equal to the minimum and maximum
intensities of nebulosity. The limiting magnitude is $\sim 22$.  }
\end{figure}

\newpage
\noindent
degree near the northern rim of the SNR. Some nebulosity is seen in this  
image, and it matches the nebulosity in the Schmidt plates.  It has,  
however, been rendered clearer than the nebulosity seen in the Schmidt  
plates by judicious use of the grey scale limits. It is not yet known if  
the nebulosity is caused by the COMPTEL/ROSAT
SNR or by other phenomena. It is hoped that, when further CCD images
become available that cover the entire SNR, the cause of the
nebulosity will become clearer.

\vspace{1ex}
\section{Discussion:}
The cosmic radiation was discovered in 1912, yet its origin has not yet  
been determined. It is generally assumed that cosmic rays with energies  
up to about $10^{15}$ eV are accelerated in the shocks of supernova  
remnants by the Fermi mechanism, but this has not been confirmed (Yock  
1998, Normille 1999). Unfortunately, the present study has also failed to  
provide confirmation. The measured flux of gamma-rays from the  
COMPTEL/ROSAT SNR is at least an order of magnitude less than the  
prediction. The apparent disagreement between observation and theory may  
be indicative of an anomaly of the supernova that caused the  
COMPTEL/ROSAT SNR, or of the interstellar medium in which it exploded  
(Chen and Gehrels 1999).

%
%
%
%
%
\vspace{1ex}
\begin{center}
{\Large\bf References}
\end{center}
\noindent
Abe, F.  et al, 1996, in "Variable Stars \& the Astrophysical Returns of   
Microlensing Surveys", ed R. Ferlet et al, Editions Frontieres, Paris (1996).    \\
Abe, F.  et al, 1999, Astronomical Journal, {\bf 118}, 261 \\
Acharya, B.S. et al, 1990, Nature {\bf 347}, 364 \\
Alcock, C. et al, 1997, ApJ  {\bf 491}, 436 \\
Allen, W. H. et al, 1993, Phys. Rev. D   {\bf 48}, 446 \\
Allen, W. H. et al, 1995, Proc. 24th ICRC {\bf 2}, 447 \\
Aschenbach, B., 1998, Nature  {\bf 396}, 141 \\
Bond, I.A. et al, 1988, Phys. Rev. Lett.  {\bf 60}, 1110 \\
Chen, W. and Gehrels, N., Astrophysical Journal (in press)        \\       
Drury, L. O'C., Aharonian, F.A., \& Volk, H. J., 1994, A \& A,  {\bf 287}, 959  
\\
Iyudin, A. et al, 1998, Nature  {\bf 396}, 142 \\
Li, T. and Ma, Y. 1983, Astrophysical Journal {\bf 272}, 317  \\                                                                        
Normille, D. 1999, Science {\bf 284}, 734 \\               
Muraki, Y. et al, 1999, Prog Theor Phys Suppl {\bf 133}, 233 \\
Naito, T. \& Takahara F., 1994, J. Phys G,  {\bf 20}, 477 \\
Reid, M. et al, 1998, Australian J Astronomy  {\bf 7}, 79 \\
Rhie, S. et al 2000, Astrophysical Journal (in press)  \\                 
Yanagisawa, T. et al, (submitted to Experimental Astronomy) \\
Yock, P., 1998, in "Black Holes \& High Energy Astrophysics", (ed. H.  
Sato and N. Sugiyama, Frontiers Science Series No. 23, Universal Academy
Press, Tokyo)\\
\end{document}